\documentclass[aip,jcp, preprint]{revtex4-1}
\usepackage{graphicx,color}
\usepackage{amsmath,amssymb,latexsym}
\usepackage[greek,english]{babel}
\usepackage{dcolumn}
\usepackage{bm}
\usepackage{float}
\usepackage[normalem]{ulem}

\usepackage{color}

\begin{document}

\title{Multiblob coarse-graining\\ for mixtures of long polymers and soft colloids}

\author{Emanuele Locatelli}
\email{emanuele.locatelli@univie.ac.at}
\affiliation{Faculty of Physics, University of Vienna, Boltzmanngasse 5, A-1090 Vienna, Austria}

\author{Barbara Capone}
\email{barbara.capone@univie.ac.at}
\affiliation{Faculty of Physics, University of Vienna, Boltzmanngasse 5, A-1090 Vienna, Austria}

\author{Christos N. Likos}
\email{christos.likos@univie.ac.at}
\affiliation{Faculty of Physics, University of Vienna, Boltzmanngasse 5, A-1090 Vienna, Austria}

\begin{abstract}
Soft nanocomposites represent  both a theoretical and an experimental  challenge due to the high number of the microscopic constituents that strongly influence the behaviour of the 
systems. An effective theoretical description of such systems invokes a reduction of the degrees of freedom to be analysed, hence requiring the introduction of an efficient, 
quantitative,  coarse-grained description. 
We here report on 
a novel  coarse graining approach based on a set of transferable potentials that quantitatively reproduces properties of mixtures of linear and star-shaped 
homopolymeric nanocomposites.  By renormalizing  groups of monomers into a  single effective potential 
between a $f$-functional star polymer and an 
homopolymer of length $N_0$, and through  a scaling argument, it will be shown how a substantial reduction  of the  to degrees of freedom allows for a full quantitative description of 
the system. Our methodology is tested upon full monomer simulations for systems of different molecular weight, proving its full predictive potential.  
\end{abstract}


\maketitle

\section{Introduction}

Over the last decades, soft matter reached a paradigmatic importance  in the scientific community. 
Novel building blocks, ranging from the nanoscale to the mesoscale have been designed; materials scientists  
have predicted and developed materials with tunable  properties, which can be 
controlled by means of simple changes in chemical or physical conditions.
These are based on building blocks that, under specific thermodynamical conditions, are able to assemble into specific target structures. 
Amongst these, the striking properties of soft nanocomposites, referred here as mixtures of soft or hard colloidal particles and linear chains in the nanoscale, suspended in a solvent, gained an important role.  
Soft, polymer-based composite materials strongly catalysed researchers' interest in 
recent years\cite{Stradner:2004,Lu:2008,Mattsson:2009,Button:2012,Truzzolillo:2013,Marzi:2015}. Their complex (chemical, physical, and topological)  composition  allows for the 
introduction a set of tunable interactions between the different components of the systems, leading to a rich panorama of observable phases and  
complex aggregation properties.\cite{Foffi:2003,Marzi:2015,Paloli:2013,Lu:2008,Watzlawek:1999} A wide range of different dynamical (rheological) 
types of behavior follows\cite{Erwin:2010,Christopoulou:2009,Vlassopoulos:2014}, opening the path for applications of technological relevance.\cite{Chu:2007,Mohanty:2012,Nojd:2013}
In this framework, star polymers have emerged as archetypal examples of versatile colloidal particles\cite{Grest:1996,Likos:2001}. From the experimental perspective, the synthesis of 
star polymers is nowadays well controlled\cite{Zhou:1993}, and hence they represent a reliable model system\cite{Grest:1996,Stiakakis:2010}. At the same time, from the theoretical 
perspective, they stand as a unique link between polymeric and colloidal physics: at low functionality $f$ and low density they resemble linear polymeric chains, while for high values 
of $f$ they can be very well described as stiff, sterically stabilized colloids. Moreover, they display a characteristic ultrasoft logarithmic effective interaction\cite{Likos:1998,Jusufi:1999}, 
which leads to very rich phase diagrams, for pure systems\cite{Watzlawek:1999,Foffi:2003,Stiakakis:2010} as well as for mixtures of either 
stars and colloidal particles\cite{Dzubiella:2001,Truzzolillo:2013,Marzi:2015} or stars and linear polymeric chains, whose radius of gyration $R_c$ is at most of the same order of 
magnitude of the average size of the stars $R_s$\cite{Camargo:2009,Camargo:2010,Stiakakis:2002,Stiakakis:2005,Lonetti:2011,Truzzolillo:2011,Wilk:2010}. Star-linear 
homopolymer nanocomposites 
are expected to show interesting properties also for size ratios $ q = R_c/R_s >1$, i.e., for polymer chains longer than the stars. However, 
due to the complexity of the theoretical description, such mixtures of stars and linear chains of arbitrary length\cite{Hooper:2006,Kumar:2010,Kumar:2013}, 
have attracted much less attention to-date.

The purpose of this paper is to introduce a novel multiscale coarse-graining framework, suitable for a quantitative description of mixtures of long homopolymeric chains and star polymers of arbitrary functionality and molecular weight.  
Previous studies that focused on the realm in which the chains were shorter than the stars\cite{Camargo:2009,Camargo:2010,Stiakakis:2002,Stiakakis:2005,Lonetti:2011,Truzzolillo:2011,Wilk:2010} employed a coarse-graining strategy in which the whole polymer chain was represented by a single degree of freedom, typically its central monomer,
in the spirit of, e.g., the paradigmatic Asakura-Oosawa model of colloid-polymer mixtures\cite{Asakura:1954}. The latter loses its validity fundamentally, not just 
quantitatively, however, when the chains
grow larger than the colloids\cite{Bolhuis:2003}. Similarly,   
to systematically analyze the properties of nanocomposites made of many monomeric units in regions of the phase space, up to the semi-dilute regime, 
in which drastic departures of the polymer conformation from an average, `soft sphere' shape are expected, 
it appears essential to develop a novel coarse graining strategy. In this approach, we employ a regrouping of degrees 
of freedom that allows for a simplification of the analysed system while retaining the ability of reproducing its essential features
as a polymer {\it chain}, i.e., as a linear succession of building blocks, irreversibly connected to one another.
To attain such a description, we here introduce a multiscale strategy that allows, on the one hand, 
to retain the complexity arising from the many body interactions between  polymeric chains and  star polymers while, on the other hand, it 
permits for a drastic reduction of the number of units used to describe the system. Hence, at the coarse-grained level, the long chain is replaced by a succession of $N_b$ effective 
monomers, or \textit{blobs}, each of which represents a small sub-part of the $N_c$-long polymeric  chain, each blob made of $N_0 = N_c/N_b$ monomers. Every blob sees the star as a single, coarse-grained object; as will be shown in the following sections, the star-blob interaction $V_{sb}(r)$ is obtained by computing, in the zero-density limit, the effective potential between a star polymer and a short polymer chain of length $N_0$,which is fixed throughout the paper to be $N_0$=10. The blob-based approach proposed here, permits a realistic description of  long polymeric  chains, accessing information on the conformational properties of the latter while reducing the 
star to a single interaction center, hence achieving a huge computational gain with respect to a full monomer resolved simulation.

In what follows, after a brief description of the numerical methods in Section \ref{sec:numerical}, 
we present in Section \ref{sec:starblob} a computational derivation and a theoretical interpretation of the zero-density star-blob potential, illuminating and 
rationalising its intriguing scaling properties. Exploiting such a scaling, we provide an approximate analytical form for the computed effective potential, 
which is then proven to be valid 
for stars of arbitrary size and functionality $f \gg$1. The formalism that will be introduced will allow for a description of nanocomposites made of stars of arbitrary size and polymers of 
arbitrary length in the dilute and the semi-dilute regimes, hence becoming a powerful methodology to be used to represent the system at different densities and ratios $q=R_s/R_c$. The multiscale coarse graining is 
then tested in Section \ref{sec:comparison}
against full monomer simulations for different size ratios, by comparing the effective potential between a star-polymer and a (long) linear chain, $V_{sc}(r)$, both computed at a 
monomer-resolved and at a  coarse-grained level. We find that the coarse graining method is able to reproduce the monomer-resolved results within an excellent 
degree of agreement. We summarize and draw our conclusions in Section \ref{sec:conclusions}, whereas some more technical aspects of the work, related to the
structural unit employed for the coarse-graining are discussed in the Appendix.

\section{Numerical methods}
\label{sec:numerical}

Monomer-resolved and coarse-grained simulations are performed employing standard Langevin Dynamics. 
The monomer size was set as unit of length, $\sigma = 1$, [see Eq.\ (\ref{eq:LJ}) below] and further the momoner mass 
$m=1$, Boltzmann's constant $k_{\rm B}=1$, and absolute temperature $T=1$ were selected to form the physical units of the system.
The friction coefficient was chosen to have the value $\gamma = 1$ and the 
Langevin equations of motion were integrated 
with an elementary timestep $\Delta t=10^{-3}$. We always consider the so-called `zero density limit', i.e., systems made of two objects, 
a star polymer and a linear chain,
as a standard procedure used to compute effective potentials. At the monomer-resolved level, we consider star polymers made of $f$ arms and $N$ monomers 
per arm ($N$ being the arm length or  the degree of polymerization of the star), grafted on a central anchoring point. 
We first compute the star-blob effective potential considering, as already mentioned, the interaction of a star polymer of functionality $f$, 
and  a short chain of length $N_0=10$ monomers. In Section \ref{sec:comparison}, we also consider longer chains, of length $N_c$. 

All the monomers in the simulations are interacting through a purely repulsive truncated and shifted Lennard-Jones potential (i.e, good solvent conditions):
\begin{equation}
 V_{mm}(r) = 
\begin{cases}
4 \epsilon 
\left[ \left(\frac{\sigma}{r} \right)^{12} -
  \left(\frac{\sigma}{r} \right)^{6} \right] 
+ \epsilon; &{\text {for}}\,\, r<2^{1/6}\,\sigma, \cr
0; &{\text {for}}\,\, r \geq 2^{1/6}\,\sigma,
\end{cases}
\label{eq:LJ}
\end{equation}
where $\epsilon=k_{\rm B} T$. Each neighbouring pair constituting the backbone of each arm of the star, as well as the backbone of the linear chains, is held together via a FENE (finite extensible nonlinear elastic) potential
{\small
\begin{equation}
V_{\rm {FENE}}(r) = 
\begin{cases}
-15 \epsilon \left( \frac{R_{\rm max}}{\sigma} \right)^2  \ln \left[ 1 - \left( \frac{r}{R_{\rm max}} \right)^2 \right] ; &{\text {for}}\,\, r \leq R_{\rm max}, \cr
\infty; &{\text {for}}\,\, r > R_{\rm max},
\end{cases}
\label{eq:FENE}
\end{equation}
}
where $R_{\rm max}$ is the maximum extension of the bond, chosen to be $R_{\rm max}=1.5\sigma$.

The natural choice of effective coordinates to describe the star and the short polymer blob are the central anchoring point of the former and the center of mass
of the latter. Accordingly, we aim at calculating the star-blof effective potential $V_{sb}(r)$, where $r$ is the distance between the two effective coordinates. 
For this purpose,  we employ the Widom insertion method\cite{Mladek:2010}, which guarantees excellent precision at both short and large distances, being at the same time 
computationally cheap and fast. We briefly describe it here: after equilibration, every ${\mathcal N}=10^4-10^5$ time steps, we fix a configuration of the system, 
making sure that the two objects considered for the calculation of the effective potential (star and blob) are not interacting. 
We then attempt to insert one object (the short chain), moving its center of mass at a fixed distance $r$ from the center of the other (the star); 
we choose thereby a random orientation for the object we are moving. For every insertion at center-to-center of mass separation $r$,
we compute the Boltzmann factor $e^{-\beta \Delta U(r)}$, where $\Delta U(r)$ is the energy difference between the original and the new configuration. 
The latter includes only cross-interaction terms between the star and the blob, since the reference state is the one of infinite separation between
the two, where they are not interacting with one another. 
This approach allows us to calculate the inter-particle radial distribution function, as 
\begin{equation}
g_{sb}(r) = \left\langle e^{-\beta \Delta U(r)} \right\rangle_{N_{\rm trials}},
\end{equation}
where the average is done over $N_{\rm trials} \approx 10^5$ trial insertions. 
We than obtain the effective inter-particle interaction potential $V_{sb}(r)$ as 
\begin{equation}
\beta V_{sb}(r) = -\ln\,g_{sb}(r), 
\label{eq:potential}
\end{equation}
where $\beta \equiv \left(k_{\rm B}T\right)^{-1}$.

At the coarse-grained level, we then consider long chains of $N_c$ microscopic monomers, which are composed by $N_b$ blobs, 
each one of them representing $N_0$ monomers, 
so that $N_c = N_0 \cdot N_b$. As already 
mentioned, the star is, in this coarse-grained level, represented as a single object, and each blob interacts with it via the star-blob potential $V_{sb}(r)$. 
Within the linear chain, each blob is connected to its neighbours by a 
harmonic potential
\begin{equation}
	\beta V_{\rm conn} (s)= \frac{\kappa}{2}(s - s_0)^2,
\end{equation}
where $s$ is the separation between the centers of mass of two (connected) blobs, $\kappa$ = 0.11 and $s_0 = R_b$, $R_b$ being the radius of gyration of a 
polymer chain of length $N_0$. Moreover, each blob interacts with any other blob, connected to it or not, through a Gaussian steric potential\cite{Pierleoni:2007}
\begin{equation}
	\beta V_{\rm steric}(s) = A \exp \left[ -\alpha(s/s_0)^2 \right],
\end{equation}
where $\alpha=1.98$, $A=2.45$. The parameters $\kappa$, $A$ and $\alpha$ have 
been determined through an optimization procedure: several combinations of values were tried, until a satisfactory agreement 
between the interaction potential of a star of functionality $f=50$, arm length $N=50$ and a linear chain of $N_c=100$ in the coarse-grained and monomer resolved simulation was 
achieved, see Section \ref{sec:comparison}. This choice has proven successful for all other combinations of stars and chains tested. 

Effective potentials that will be used throughout this work  for the coarse-grained description of the homopolymeric 
chain  are computed via an iterative procedure for blobs that contain $N$=10 monomers. 
The intramolecular potentials  used to describe the chain are extracted in proximity of a  star polymer; in such a way  
both the intramolecular  and the chain-star many-body contribution are included in the effective interactions.
The fact that blobs only contain a few monomers each (hence the subsegments of the chain are off-scaling) 
and the influence on the effective representation due to the presence of the  
star,  leads to a discrepancy between the here used potentials and the ones known to well describe 
homopolymeric chains\cite{Pierleoni:2007}; effective potentials have in fact, in the 
present  work, a shorter ranged blob-blob interaction and a weaker tethering. Such a discrepacy arises,  
as depicted in Fig.~\ref{fig:snapshot}, from the stretching that the chain acts 
when the centre of mass of the homopolymer approaches the centre of the star. 
To mimic such a conformational property, the intramolecular effective potentials computed within the star 
are softer than expected. Moreover, following previous results~\cite{Poier:2015}, the strength 
of the blob-blob interaction is reduced with respect to its value in the absence of the star, again 
as a consequence of the stretching of 
the chain in the star interior.

Finally, concerning the computation of the effective $V_{sc}(r)$ potentials between a star center and 
the center of mass of an arbitrarily long linear chain, we have performed standard Umbrella Sampling Molecular Dynamics simulations 
in both the coarse-grained and monomer-resolved description. Briefly, we added a bias force in the system in order to improve the sampling of the configurational space, 
dividing the domain of separations $r$ between the star center and the center of mass of the long chain into `sampling windows'. 
A standard choice in Langevin Dynamics simulations is given by adding a harmonic spring with a force $f_{\rm bias}^{j}(r)$ given by
\begin{equation}
\beta f_{\rm {bias}}^{j}(r)= -\kappa_{\rm {bias}} (r - r_{0,j}),
\end{equation}  
The parameters $\kappa_{\rm bias}$ and $r_{0,j}$ set the width and the center of the sampling windows, respectively. 
The distance between the centres of neighbouring windows is $\Delta r_{\rm bias} = r_{0,j+1} - r_{0,j}$. 
We sample different windows in parallel, separate simulations, setting for each one a different value of $r_{0,j}$, and we compute the biased radial distribution function 
$\tilde{g}_j(r)$ in each window. We allow for an overlap between neighbouring windows, setting $\Delta r_{\rm bias} = \sigma$ and $\kappa_{\rm bias}=1$. 
Thus, employing a standard unbiasing procedure, we merge at the end the computed effective interaction potential $V_{sc}(r)$ by patching together the pieces 
from the different windows. To this end, we add together pieces $V_{sc}^{j}(r)$ from each window, 
in which the bias is removed and a vertical shift has been performed as: 
\begin{equation}
\beta V_{sc}^{j}(r) = - \ln\,\tilde{g}_j (r) - \beta V_{\rm bias}^j(r) +c_j,
\end{equation}
where $\beta V_{\rm bias}^j(r) = \kappa_{\rm bias}(r - r_{0,j})^2/2$ is the bias potential and 
the shifting constants $c_j$ are introduced to match the pieces of the effective potentials obtained in neighbouring windows.

\section{Star-blob effective interaction}  
\label{sec:starblob}

We now present the results obtained for the effective interaction between a short linear chain and star polymer and introduce the proposed generalizable multiscale methodology used 
to reproduce the effective interaction between star polymers of arbitrary functionality and linear chains of arbitrary length. For this purpose, the effective interaction between a star 
polymer of functionality $f$ and a linear chain made of $N_0=10$ monomers has been first computed. This will allow us to extract an effective star-blob potential $V_{sb}(r)$ for a 
fixed $N_0$, which will be then used as a building block to represent, within an effective interaction framework, systems made of star polymers of functionality $f$ and linear 
chains of length $N_c \gg N_0$. We here report the numerical results regarding the star-blob effective interaction. 
The star-blob effective potentials have been obtained numerically 
as described in Section~\ref{sec:numerical}. 

\begin{figure}[!h]
\begin{center}
\includegraphics[width=0.65\textwidth]{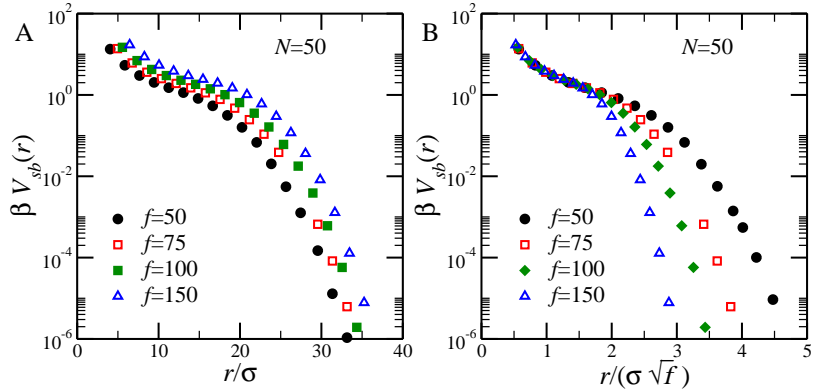}
\caption{(a) Semi-log plot of star-blob interaction potential, obtained from 
numerical simulations, as function of $r/\sigma$ for star polymers of different functionality $f$, 
fixed arm length $N=50$ and a short chain of length $N_0=10$. (b) Semi-log plot of the same data as in panel (a), as function of $r/\sigma \sqrt{f}$.}
\label{fig:shortchain_f}
\end{center}
\end{figure}

In Fig.~\ref{fig:shortchain_f}(a) we show the interaction potentials as function of $r/\sigma$ for star polymers of different functionality $f$, but same arm length $N=50$, and the short 
chain of length $N_0 = 10$; it can be seen that, as expected, the potentials get significantly more repulsive for increasing $f$ at fixed $r/\sigma$. 
Fig.~\ref{fig:shortchain_f}(b) demonstrates an interesting scaling property. If we plot the interaction potentials as function of $r/(\sigma \sqrt{f})$, a nice collapse on a unique curve 
takes place, for relatively small values of $r/(\sigma \sqrt{f})$, roughly up to $R_{sb}/(\sigma \sqrt{f})$,  where $R_{sb} = R_s +R_b$, 
$R_s$ being the gyration radius of the star and $R_b=1.548\,\sigma$ being the radius of 
gyration of the $N_0 = 10$-short chain. 
For values larger than $r/\sigma \sqrt{f}$, the aforementioned scaling is not valid, thus rendering the overall size of the star the important length scale.

Such a result opens the path for a scaling theoretical prediction of the shape of the potential. The small $r$ collapse of the effective potential can be interpreted using the Daoud and 
Cotton representation of a star polymer\cite{Daoud:1983}: for 
distances 
$r < R_s$ from its center, the star can be modelled as a succession of concentric shell of Daoud-Cotton (DC) blobs of size $\xi(r)$; 
the local correlation length $\xi(r)$ 
increases as we go further out from the center of the star as
\begin{equation}
\xi (r) \propto r /\sqrt{f},
\label{eq:xiofr}
\end{equation}
as long as the blobs are within a distance, from the anchoring point, comparable with the radius of gyration $R_s$. 
Within the star interior, an inhomogeneous semidilute polymer solution is formed and the DC blob is the local correlation length
$\xi(r)$. Accordingly, and in full analogy with colloid-semidilute polymer interactions\cite{Sear:1997}, 
the potential between the short chain and the semidilute-solution DC blobs is the same for DC blobs of the same size $\xi(r)$, 
irrespectively of where the latter have their
physical origin.
Consider therefore two stars of different functionalities, $f_1$ and $f_2$, and take the interaction of the small chain with a DC blob of size $\xi(r)$ in the first star of functionality $f_1$. 
Following the Daoud-Cotton model, a DC blob of the same size $\xi(r)$ in a star of functionality $f_2$ will be found at a distance $r \sqrt{f_2/f_1}$ with respect to the center of the 
second star. Accordingly, the free energy cost of insertion of the short blob in the interior of the two stars fulfills the property
\begin{equation}
V_{sb}(r, f_1) = V_{sb} \left( r \sqrt{\frac{f_2}{f_1}}, f_2 \right)
\end{equation}    
where we explicitly inserted the $f$-dependence in the interaction potential. The implication is now a scaling property of the latter with $r$ and $f$, namely
\begin{equation}
V_{sb}(r,f) = \phi(r/\sqrt{f}),\qquad\qquad (r \lesssim R_s)
\end{equation}
with some function $\phi(x)$.
Since this must hold for any $f$, we find that a universal function $\phi(r/\sqrt{f})$ exists for such potentials, as long as the blob model holds. In other words, since the short chain is a 
small object (its radius of gyration is much smaller than that of the whole star), 
it is able to probe the internal structure of the star. It is thus reasonable to find a signature of the properties of the monomer distribution within the star in the inter-particle potential, as 
a consequence of the exploration done by the small probe. Finally, we stress that this argument is valid as long as the object interacting with the star polymer is sufficiently small with 
respect to $R_s$.

\begin{figure}[!h]
\begin{center}
\includegraphics[width=0.65\textwidth]{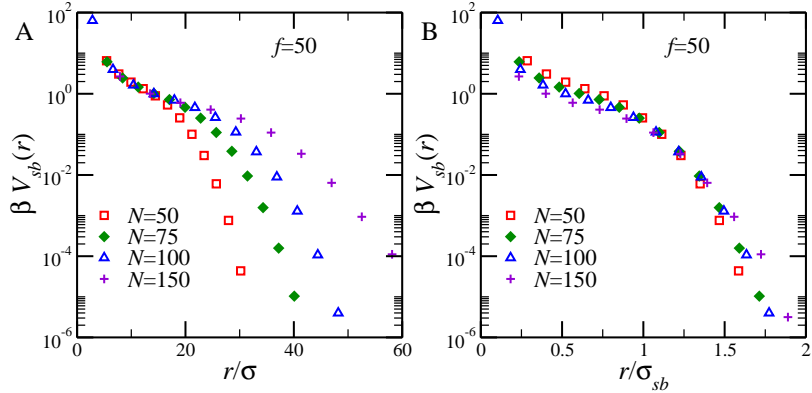}
\caption{(a) Semi-logarithmic plot of star polymer - short chain interaction potential, 
obtained from numerical simulations, as function of $r/\sigma$ for star polymers of different 
arm length $N$, fixed functionality $f=50$ and a short chain of length $N_0=10$. (b) Semi-log plot of the same 
data in (a), as function of $r/\sigma_{sb}$, $\sigma_{sb}=4(R_s + R_b)/3$.}
\label{fig:shortchain_N}
\end{center}
\end{figure}

To interpret the large $r$ behavior, we now consider the interaction potentials as function of $r/\sigma$ for star polymers of different arm length $N$, fixed functionality $f=50$ and 
the short chain, see Fig.~\ref{fig:shortchain_N}(a). We observe that for small values of $r/\sigma$, the potentials do not appreciably depend on $N$, which is another
manifestation of the fact that, as long as we are not too far away from the star center and thus the star is a semidilute polymer solution, only the local
correlation length $\xi(r)$ matters. The latter is expressed by Eq.\ (\ref{eq:xiofr}) above, so that it is the same for fixed $r$ for fixed $f$ as long as $r < R_s$.
Thus, the arm length $N$ at fixed 
functionality affects the potential only at large $r/\sigma$, extending the range of the potential for increasing $N$. In Fig.~\ref{fig:shortchain_N}(b), 
we plot the same data as function of 
$r/\sigma_{sb}$, where\cite{Mayer:2007} $\sigma_{sb}=4(R_s + R_b)/3$. 
We now find the signature of a common scaling for values $r/\sigma_{sb} \gtrsim 1$, which is in full agreement with previous findings
in which the central monomers of both the star and the chain were employed as effective degrees of freedom\cite{Mayer:2007}. Such features lead to a general 
prediction of the shape of the interaction potential for arbitrary $f$ and $N$ and can be exploited to obtain an approximate, analytical form of the star-blob interaction potential. All 
functionalities tested produce potentials that are well reproduced by the theoretical approach, thus allowing to extrapolate effective potentials for stars with $f \gg$ 1, comparable to 
those used in experimental realizations of similar systems\citep{Stiakakis:2002,Stiakakis:2010,Truzzolillo:2013}. Note that we only consider large stars, 
such that the scaling $R_s \sim f^{1/5}N^{\nu}$, with the Flory exponent $\nu = 0.588$ holds. 

\begin{figure}[!h]
\begin{center}
\includegraphics[width=0.65\textwidth]{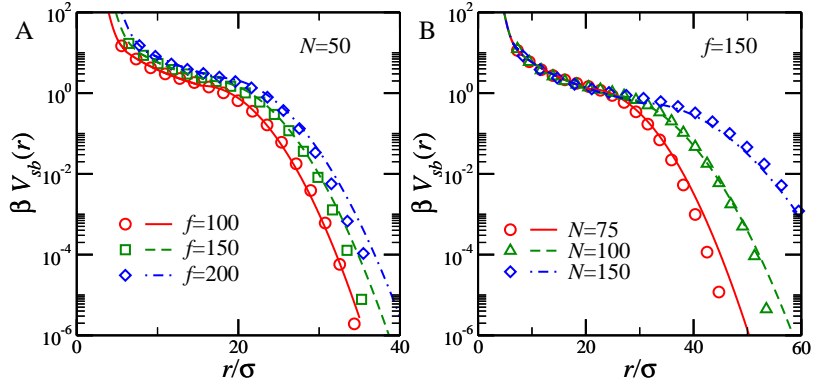}
\caption{Comparison between Eq.~(\ref{eq:analytic}) (lines) and numerical (open symbols) star polymer - short chain ($N_0=10$) 
interaction potentials, as function of $r/\sigma$. (a)Star polymers of different functionalities $f$, 
fixed arm length $N=50$; (b) Star polymers of fixed functionality $f=150$ and different arm length $N$.}
\label{fig:pot_f2_scal}
\end{center}
\end{figure}

The collapse of the data reported in Fig.~\ref{fig:shortchain_f}(b) highlights two distinct power-law regimes, in the region $r<R_{sb} = R_s + R_b$. 
The first regime, valid up to $r/\sigma \approx \sqrt{f/e}$ is characterized by a very steep descent of the potential. 
The second regime, valid for  $\sqrt{f/e} \leq r/\sigma \leq R_{sb}$, shows instead a slower decay. Last, for $r/\sigma > R_{sb}$, 
we employ a Gaussian function, as suggested by comparison with results obtained previously in the literature\cite{Mayer:2007}.
We find thereby a general analytical expression (data fit) for $\beta V_{sb}(r)$ in the form:
\begin{widetext}
\begin{equation}
\beta V_{sb}(r) = 
\begin{cases}
v_1 \left( \frac{r}{\sqrt{f}} \right)^{-b_1} + c_1; & \qquad \text{for } r/\sigma \leq \sqrt{f/e}, \cr 
v_2 \left( \frac{r}{\sqrt{f}} \right)^{-b_2};   & \qquad \text{for } \sqrt{f/e} < r/\sigma \leq R_{sb}, \cr 
v_3 \exp \left[-(r-R_{sb})^2/(w R_{sb}^2) \right]; & \qquad \text{for } R_{sb} < r/\sigma, 
\end{cases}
\label{eq:analytic}
\end{equation}
\end{widetext}
where $v_1=0.0477$, $b_1=8.1279$, $c_1=9.0306$, $v_2=3.7973$, $b_2=2.091$, have been obtained through a  fitting procedure, bound to ensure that the potential is continuous 
at $r/ (\sigma \sqrt{f}) = 1/\sqrt{e}$. The parameter $v_3$ has to be chosen such that the potential is continuous at $r = R_{sb}$. Last, the parameter $w$, which regulates the width of 
the Gaussian, is obtained again by a fitting procedure. In Figs.~\ref{fig:pot_f2_scal}(a) and \ref{fig:pot_f2_scal}(b), we report the comparison between numerical results and 
Eq.~(\ref{eq:analytic}). In Fig.~\ref{fig:pot_f2_scal}(a) we consider stars of fixed arm length $N=50$ and functionality $100<f<200$: fixing $w=0.12$ provides an excellent consistent 
agreement between the numerical and analytical star-blob potential for the chosen $N$. For stars with longer arms, fixing $w=0.145$ yields consistently an excellent agreement with 
the numerical data, see Fig.~\ref{fig:pot_f2_scal}(b). We remark that in both cases the agreement is excellent: the methodology proposed provides a closed, precise analytical 
description of $V_{sb}(r)$.  

\section{Validation of the coarse-graining approach}
\label{sec:comparison}

In order to validate the effective potentials and multiscale methodology proposed in this paper, we performed a series of simulations 
-- both within a full-monomer and the coarse-
grained realization -- to extrapolate effective potentials $V_{sc}(r)$ between star polymers of arbitrary $f$ and linear chains of arbitrary $N_c$. In order to simulate 
the hybrid star-chain system
within the coarse-
grained multiscale approach, we start from the effective potentials described in Eq.~(\ref{eq:analytic}). Arbitrary combinations ($N_c$, $f$) can be chosen 
by simply varying the two values $f$ and $R_{sc} = (R_s +R_c) \sim f^{1/5}N^{\nu} + N_c^{\nu}$. The choice of a given $N_c$ will instead be reflected on the number of blobs (beads) 
used to represent the homopolymeric chain as  $N_c = N_b \cdot N_0$.
 
\begin{figure}[!h]
\begin{center}
\includegraphics[width=0.65\textwidth]{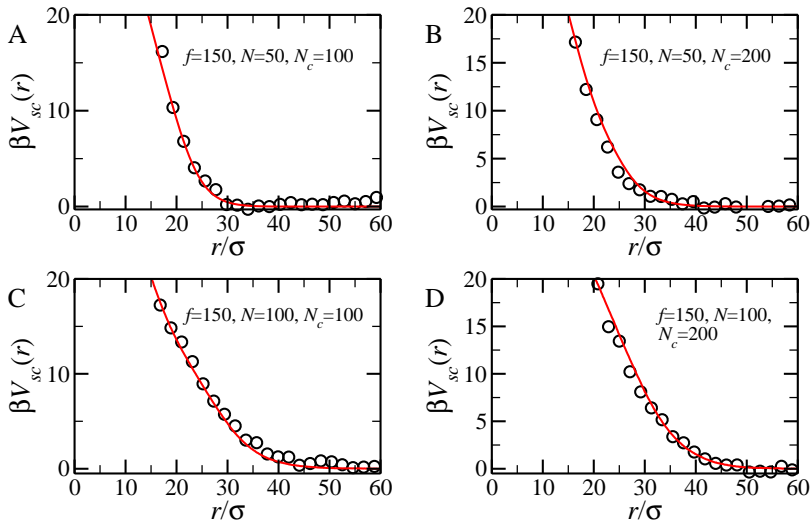}
\caption{Comparison between full monomer (open black dots) and coarse grained (red lines) star-long chain interaction potential, computed with respect to the center of mass of the 
chain, for star polymers of fixed functionality $f=150$ and (a) $N=50$, $N_c=100$; (b) $N=50$, $N_c=200$; (c) $N=100$, $N_c=100$; (d) $N=100$, $N_c=200$.}
\label{fig:compchain_50}
\end{center}
\end{figure}

The striking comparison between effective potentials obtained within the two representations is reported in Fig.~\ref{fig:compchain_50}. Here, results obtained for fixed $f=$150, two 
different arm length (namely, $N=$50, 100) and two different chain lengths ($N_c=$ 100, 200) are reported. 
We want to stress that the potentials obtained within the coarse grained representation show an excellent agreement with the monomer resolved ones, while the drastic reduction of 
the degrees of freedom grants an impressive computational gain. We can measure the computational gain as follows: we run both simulations for the same amount of time $\Delta t$ on the same machine, and we compare the number of MD steps performed within such a time interval. We observe a computational gain of two orders of magnitude: the coarse grained simulations are roughly 350-250 times faster than the full monomer ones.\\
\begin{figure}[!h]
\centering
\includegraphics[width=0.65\textwidth]{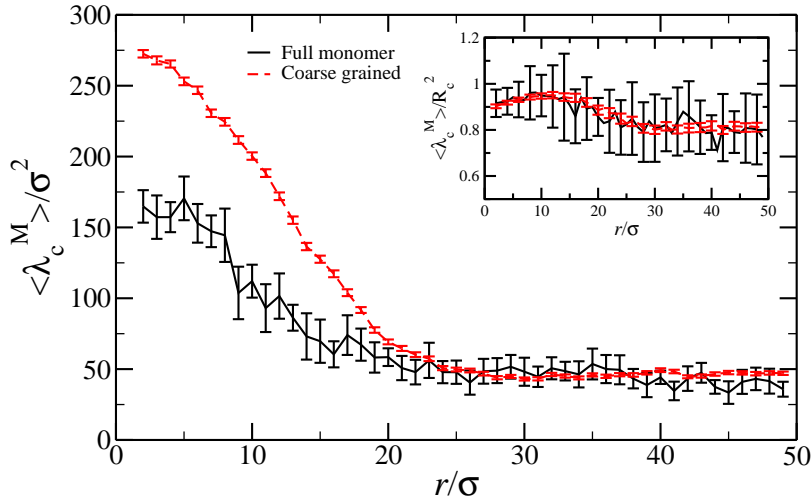}
\caption{Main panel: time-averaged largest eigenvalue of the gyration tensor of the linear chain $\langle \lambda_c^M \rangle$ as function of the distance $r$ between the center of the star and the center of mass of the linear chain, in both full monomer (black full line) and coarse grained (red dashed line) representations. Same MD units are used in both cases. Inset: ratio between the largest eigenvalue and the square of the radius of gyration, as function of the distance $r$ between the center of the star and the center of mass of the linear chain, in both full monomer and coarse grained representations (colors as in main panel).}
\label{fig:answer1}
\end{figure}
Additional, strong corroboration of the validity of our approach come from the analysis of the shapes and of the orientations of the chain with respect to the vector $\vec{r}$, connecting the center of the star and the center of mass of the chain.
We first analyze the shapes of the chain, by computing the gyration tensor. In Fig.~\ref{fig:answer1}, we report the time-averaged largest eigenvalue of the gyration tensor of the linear chain $\langle \lambda_c^M \rangle$ as function of the distance $r$ between the center of the star and the center of mass of the linear chain. In both full monomer and coarse grained representations, the largest eigenvalue is maximal when the separation between the center of the two objects is small and decreases until it reaches a constant value as the separation becomes larger and larger. Within the full monomer representation, the chain, while in proximity of the center of the star, stretches to adapt to the complex structure of the star. This is well captured by the coarse-graining model, where we replace the star structure with a radially symmetric effective potential. We can compare the two representations considering the ratio between the largest eigenvalue and the radius of gyration squared. We report this comparison in the inset of Fig.~\ref{fig:answer1}. We observe now an excellent agreement between the two results, thus confirming the capability of our coarse-graining scheme to retain important conformational information of the linear chain at any separation.\\   

\begin{figure}[!h]
\centering
\includegraphics[width=0.65\textwidth]{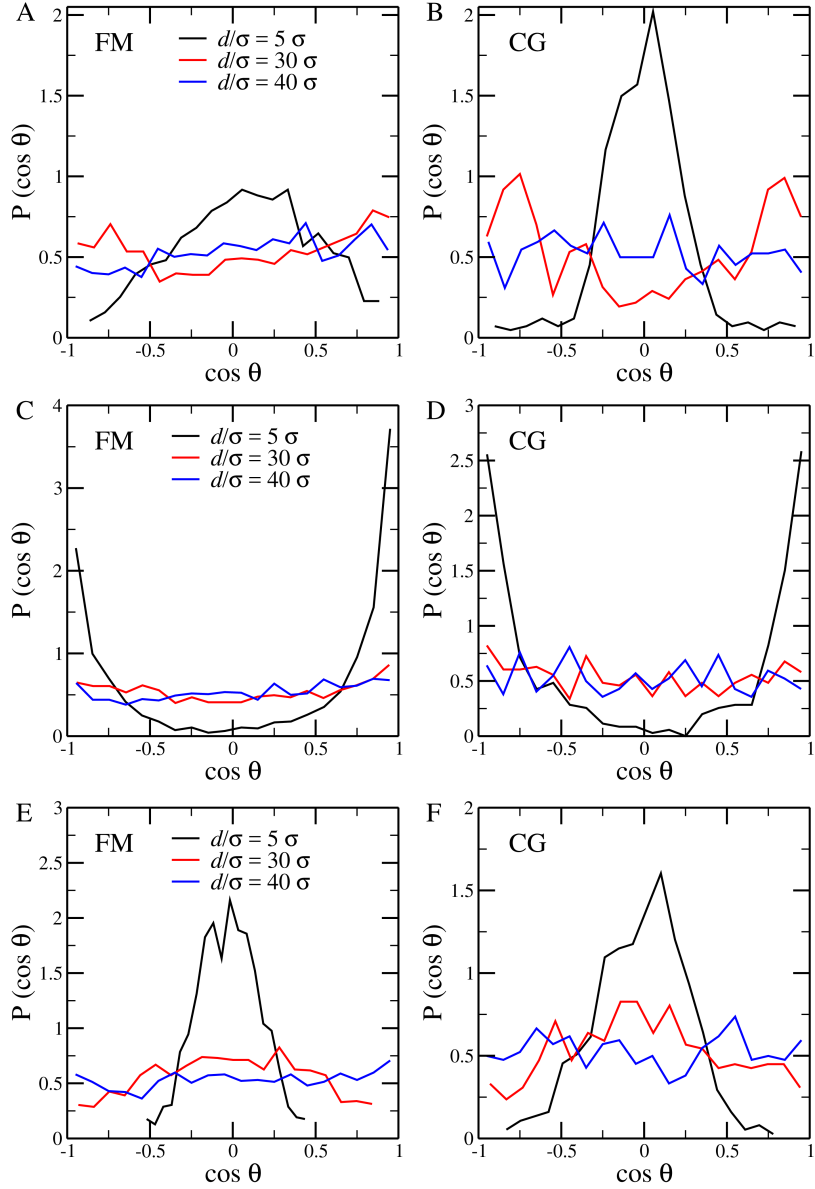}
\caption{Distribution of the cosine of the angle $\theta$ between the vector connecting the center of the star and the center of mass of the chain and the eigenvector associated to the (A,B) largest, (C,D) median, (E,F) shortest eigenvalue of the gyration tensor of the chain in both (A,C,E) full monomer and (B,D,F) coarse-grained representations. Distributions are computed for different distances $r/\sigma$ between the center of the star and the center of mass of the linear chain.}
\label{fig:answer2}
\end{figure}

We also look, as mentioned, to the relative orientation, denoted by $\theta$, between the eigenvectors corresponding to every eigenvalue, from the largest to the smallest, and the vector $\vec{r}$ connecting the star and chain centers. In the different panels of Fig.~\ref{fig:answer2}, we report the distribution of $\cos \theta$ for the different eigenvectors, in both full monomer and coarse-grained representations, at different separations. We notice, in all cases, a very good agreement between the two representations. For the largest and smallest eigenvalues, at short separations the distribution is peaked around $\cos \theta =$ 0; for the other eigenvalue, the distribution is peaked around $\cos \theta = \pm$ 1. These distributions tells us that, in both cases, the directions of the major and minor axis of the polymer are favored to be perpendicular to the vector connecting the star and chain centers, whereas the other one is favored to be parallel. This is consistent with the following picture: the chain is, in both representations, trying to stretch and to fit within the star at short separations, attempting at the same time to 'embrace' it, as much as possible. At large separations, all distributions becomes flat, as expected.\\This analysis shows that the multi-blob-based approach is, as a matter of fact, able to reproduce quite faithfully the conformational properties of the chain.\\

\begin{figure}[!h]
\begin{center}
\includegraphics[width=0.45\textwidth, angle = -90]{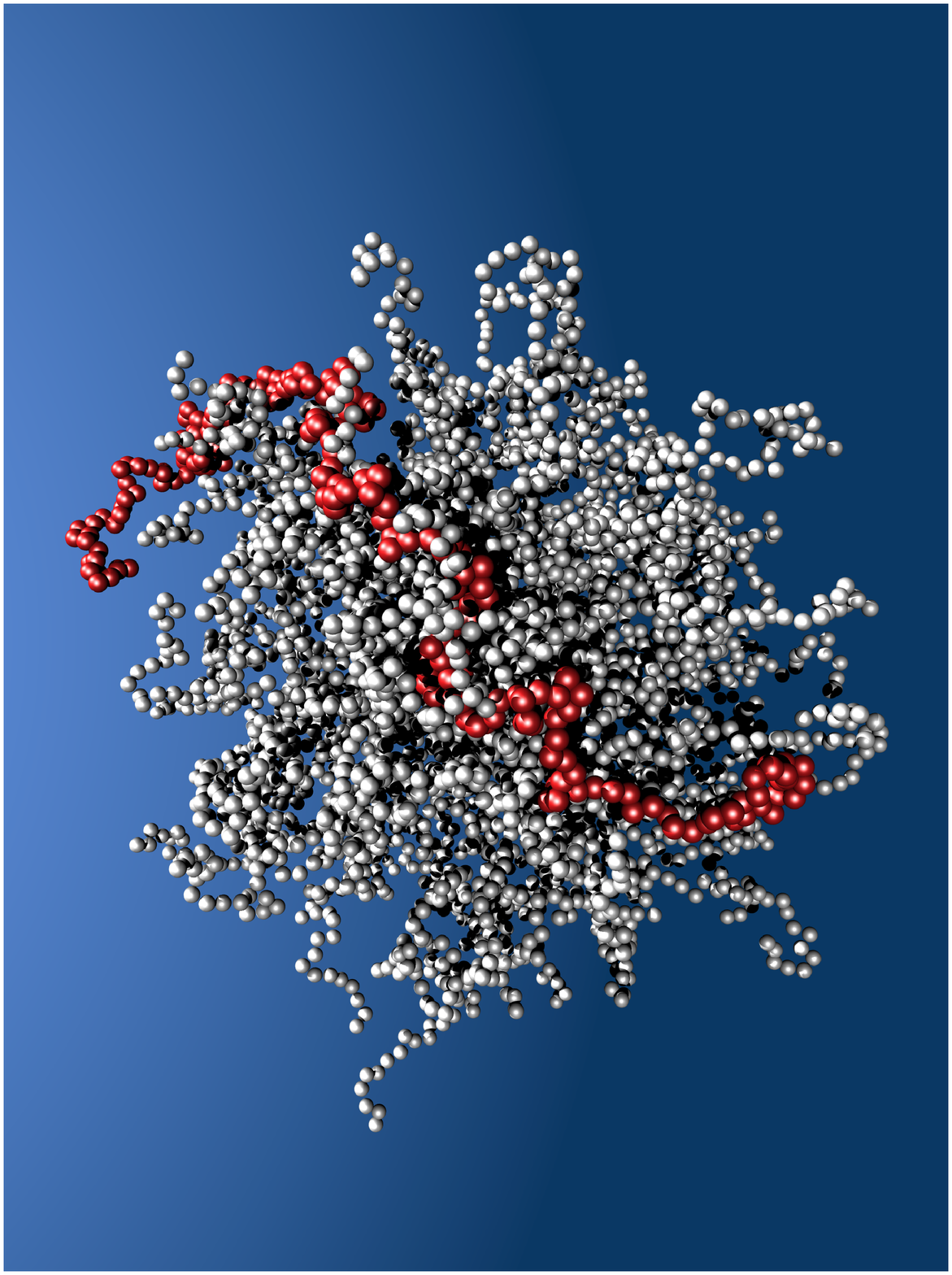}
\caption{Simulation snapshot of a star polymer with $f = 100$ arms with $N = 50$ monomers per arm, interacting with a linear
chain of $N_c = 200$ monomers, at a distance between the anchoring point of the former and the center of mass of the latter
$r = $7 $\sigma$ , corresponding to about $50\%$ of the star size.
The monomers of the stars are rendered as grey and those of the chain as red spheres.}
\label{fig:snapshot}
\end{center}
\end{figure}

Consider now a slightly different effective interaction, between the star polymer and the linear chain, namely the potential $\bar V_{sc}(r;f)$ between the
anchoring point of the star with $f$-arms and the central monomer of the chain,
where we explicitly denote the dependence on $f$ to emphasize that we are interested in a (scaling) law
to express this dependence. Evidently, the functions $V_{sc}(r;f)$ and $\bar V_{sc}(r;f)$ are different, since the
effective potentials do depend on the choice of the effective coordinates\cite{schmidt:2002}. Nevertheless, the two quantities are also related, as they both represent
a coarse-grained attempt to capture the correlations between the two macromolecules. We are interested in taking advantage of this relation
with the goal of predicting the dependence of $V_{sc}(r;f)$ on the functionality $f$ of the star. For this purpose, it is useful to note that a linear
chain with its central monomer as effective coordinate is nothing else but a star polymer with two arms. The strength of the effective interaction
between the anchoring points of two stars of functionalities $f_1$ and $f_2$ is predicted by scaling theory\cite{ferber:2000} to be determined by
the factor
\begin{equation}
\Theta_{f_1,f_2} = \left[ \left(f_1+ f_2\right)^{3/2}- \left(f_1^{3/2} + f_2^{3/2} \right) \right].
\label{thetaf1f2:eq}
\end{equation}
Setting $f_1 = f \gg1$, $f_2 = 2$ and making a Taylor expansion of the above expression up to linear order in the small parameter $2/f$,
we obtain the scaling
\begin{equation}
\Theta_{f,2} \cong 3\sqrt{f}.
\label{thetaf2:eq}
\end{equation}
Accordingly, the effective interaction $\bar V_{sc}(r;f)$ scales as
\begin{equation}
\bar V_{sc}(r;f) \cong \sqrt{f}\psi(r),
\label{sqrtfa:eq}
\end{equation}
with some function $\psi(r)$, a prediction explicitly confirmed by computer simulations for separations smaller than the typical size of the macromolecules\cite{Mayer:2007}.
The underlying physical reason for this scaling lies in the fact that for such separations, the combination of a $f$-arm star and a linear chain held at its central monomer
resembles a star with a total of $f+2$ arms\cite{ferber:2000}. Note that in comparing the theoretical predictions with simulation, the core size $R_0$ of the star
has to be subtracted, since it has a finite value in the simulation but it vanishes in scaling theory. Nevertheless, since $R_s \gg R_0$ for $N \gg 1$, this is only
a small correction\cite{ferber:2000,Mayer:2007}.

We now consider the typical, physical configurations of polymer chains when their center of mass lies at a separation $r$ from the star center which is smaller
than the star size; this is relevant for the effective potential $V_{sc}(r;f)$. In Fig.\ \ref{fig:snapshot} we show a characteristic snapshot of a linear chain with $N_c = 200$
monomers, whose center of mass is kept fixed at a distance $r = 7\,\sigma$  from the center of a multiarm star. It can be seen that the linear polymer
assumes a stretched configuration, very much akin to that of the arm of the high functionality star. This is very similar to what happens to a chain when
its central monomer is kept fixed at a distance deep inside the star interior. Moreover, the stretched configuration of the chain implies that the location
of the center of mass is not too different from that of the location of the central monomer. We surmise, therefore, that the $f$-scaling of the function
$V_{sc}(r;f)$ is the same as that of the function $\bar V_{sc}(r;f)$ given in Eq.\ (\ref{sqrtfa:eq}) above, albeit with a different function describing its
$r$-dependence, namely
\begin{equation}
V_{sc}(r;f) \cong \sqrt{f}\chi(r).
\label{sqrtfb:eq}
\end{equation}

\begin{figure}[!h]
\begin{center}
\includegraphics[width=0.65\textwidth]{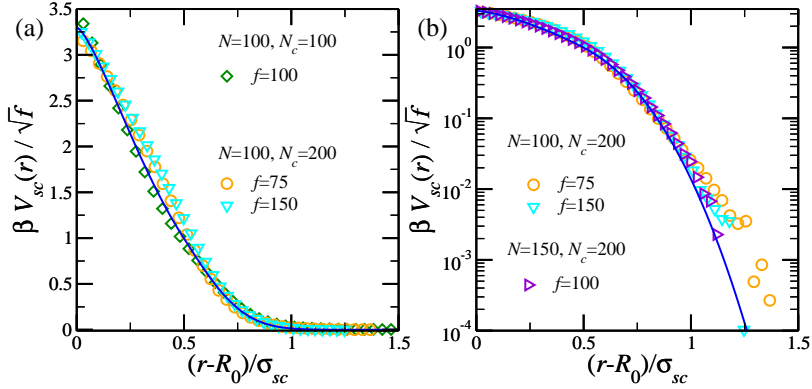}
\caption{Coarse grained star polymer - long chain interaction potential $\beta V(r)/\sqrt{f}$ as function $r/\sigma_{sc}$. Symbols are data from numerical simulations, lines are 
Eq.~(\ref{eq:fit2}). (a) Star polymers of different functionality, fixed arm length $N$=100, and chains of different arm length: $N_c$=100, 200. (b) Star polymers of different 
functionality, different arm length $N$=100, 150, and a chain of length: $N_c$=200.} 
\label{fig:scale_longchain}
\end{center}
\end{figure}

To put the prediction of Eq.\ (\ref{sqrtfb:eq}) to a test against the simulations, we should subtract from the distance $r$ the star core size $R_0$ and
choose a length scale for the star-chain separation. Following the example of the anchoring point-central monomer representation\cite{Mayer:2007},
we thus test whether the simulation data for $\beta V_{sc}(r)$ for various parameter combinations can be expressed in the form
\begin{equation}
\frac{\beta V_{sc}(r)}{\sqrt{f}} = g \left(\frac{r-R_0}{\sigma_{sc}} \right),
\label{simul:eq}
\end{equation}
with some function $g(x)$,
where $\sigma_{sc}$ is proportional to the sum of the two gyration radii, $\sigma_{sc}=4(R_s + R_c)/3$, $R_c$ being the gyration 
radius of the long chain. The value of the core radius $R_0$ for $f = 50$ has been taken from the literature, 
whereas for other values of $f$ the scaling $R_0 \sim \sqrt{f}$ has been used\cite{Grest:1996}. 
In Fig.~\ref{fig:scale_longchain}(a), the interaction potentials for stars of different functionalities $f=75, 100$, and $150$ with same arm 
length $N=100$ and chains of different lengths $N_c=100$ and $200$ are plotted. We see that, in fact, the data collapse on a master curve. We observe the same collapse 
in Fig.~\ref{fig:scale_longchain}(b), where we plot the interaction potentials for stars of different functionality $f$=75, 100, 150, arm length $N$=100, 150 and  a chains of length 
$N_c=$200. On the other hand, we stress that the collapse is valid as well in the case $(r - R_0) / \sigma_{sc} \approx$ 1: this sets the interaction range to $\sigma_{sc}$, as expected\cite{Mayer:2007}.

Using a standard fitting procedure, we find that the scaling function $g$, describing the star-long chain effective potential, can be approximated with a very good precision as 
\begin{equation}
g(x) = 
\begin{cases}
\gamma_1 \exp \left[ -(x/\alpha_1)^{\beta_1} \right] + \delta_1;  & \qquad \text{for } x < 0.5, \cr 
\gamma_2 \exp \left[ -(x/\alpha_2)^{\beta_2} \right]; & \qquad \text{for } x \geq 0.5, 
\end{cases}
\label{eq:fit2}
\end{equation}
where $\gamma_1=3.504$, $\alpha_1=0.476$, $\beta_1= 1.410$, $\delta_1=-0.204$, $\gamma_2=1.822$, $\alpha_2=0.591$, and $\beta_2=3.014$. The values chosen ensure that 
such analytical potential and its derivative are continuous. The analytical approximation Eq.~(\ref{eq:fit2}) is reported as well in Figs.~\ref{fig:scale_longchain}(a) 
and \ref{fig:scale_longchain}(b). We observe that Eq.~(\ref{eq:fit2}) is radically different from the star-blob approximation Eq.~(\ref{eq:potential}). This is another signature of the fact 
that there is a limit on the blob size, after which our coarse-graining scheme ceases to be valid.

\section{Conclusions}
\label{sec:conclusions}
We have reported in this work a novel multi-blob coarse-graining approach suitable for a quantitative description of mixtures of soft nano-composites, made of star polymers and 
linear homopolymeric chains of arbitrary functionality and molecular weight. The coarse graining is based on the computation of simple, soft transferable effective potentials between a 
star polymer and a short linear chain and the development of a theoretical framework that allows to build up a combination of the computed potentials to reproduce properties of star-
chain systems for every combination of $f$, $N$ and $N_c$. 
  
The excellent performance of the proposed multiscale approach, together with the vast reduction of the degrees of freedom  grants a remarkable computational gain and a massive 
simplification of the topology of the system without any loss in accuracy, thus becoming a novel accurate powerful tool for the study of the aforementioned nanocomposites, in the dilute and semi-dilute regimes. It is worth 
noting that our choice of $N_0$ is not completely arbitrary, but motivated from the goal of retaining as much details as possible on the chain conformation.  Nevertheless, the excellent 
results showed in the text certainly motivate us in using this approach in such conditions. Furthermore, the analytical form for the star-blob interaction we introduced, provide a 
remarkably precise operative replacement for the effective potentials computed numerically. Eq.~(\ref{eq:analytic}) allows to perform coarse-grained simulations of stars of very high 
functionality and size, very close to the ones typically used in experiments. The methodology proposed in this work will, for the first time, allow for the complete description of the star 
polymer/linear homopolymer mixtures for every $R_s/R_c$ ratio, by means of a simple construction based on one set of transferable effective potentials. It will thus be possible to 
widely explore the phase diagram of such nanocomposites in the dilute and semi-dilute regimes, with a direct link to the real experimental systems. Finally, the proposed coarse-graining strategy opens up the way for efficient modeling of concentrated solutions of soft colloids and long chain deep into the semi-dilute regime, in which nontrivial phenomena, as dynamical arrest and the emergence of strong correlations between the constituent of the system, are expected.

\section*{Acknowledgments}

We thank Dimitris Vlassopoulos and Domenico Truzzolillo for helpful discussions. B.\ C.\ acknowledges financial support through an
APART-Fellowship of the Austrian Academy of Sciences ({\"O}AW), Grant No.\ 11723.

\section{Appendix: Star-monomer effective interaction}  
\label{sec:starmon}

We report shortly here an interesting, although unsuccessful, coarse-graining scheme for star-chain mixtures. This alternative approach aim to retain the microscopic details of the 
chain, while considering the star as a single, coarse-grained object. In other words, at the coarse-grained level the beads of the chain interact with the star through an effective star-
monomer potential $V_{sm}(r)$. This star-monomer effective interaction has been calculated numerically as described in Section~\ref{sec:numerical}. 
\begin{figure}[!h]
\begin{center}
\includegraphics[width=0.65\textwidth]{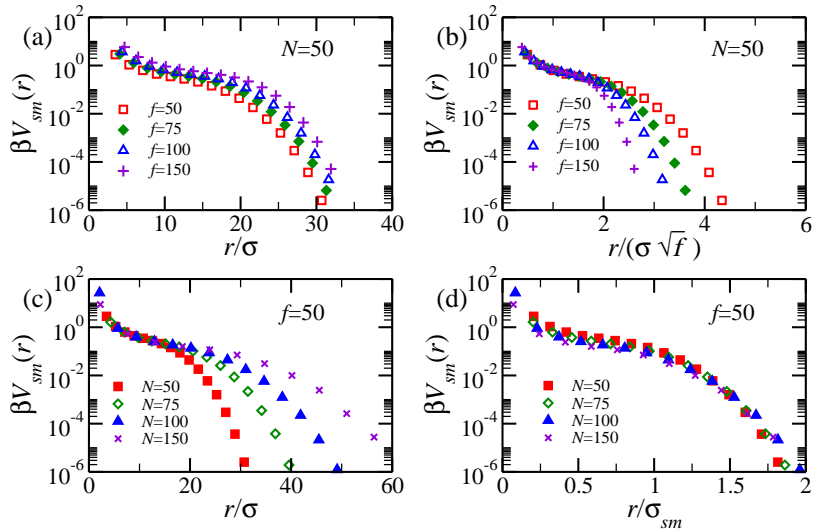}
\caption{(a) Semi-log plot of star-monomer interaction potential, as function of $r/\sigma$ for star polymers of different functionality $f$ and fixed arm length $N=50$. (b) Semi-log plot 
of the same data in left panel, as function of $r/\sigma \sqrt{f}$. (c) Semi-log plot of star-monomer interaction potential, as function of $r/\sigma$ for star polymers of different arm 
length $N$ and fixed functionality $f=50$. (d) Semi-log plot of the same data in left panel, as function of $r/\sigma_{sm}$.}
\label{fig:pot_mon}
\end{center}
\end{figure}
Similarly to $V_{sb}(r)$, we compute the interaction for stars of different functionalities and different arm lengths; results are reported in Fig.~\ref{fig:pot_mon}. 
It can be seen that in both Figs.~\ref{fig:pot_mon})(a) and \ref{fig:pot_mon}(c), the curves are very similar to the results shown 
in Figs.~\ref{fig:shortchain_f}(a) and \ref{fig:shortchain_N}(a). We have also performed the same analysis done with the star-blob effective interaction; we report the results 
in Figs.~\ref{fig:pot_mon}(b) and \ref{fig:pot_mon}(d). Again, the results are very similar to those reported 
in Figs.~\ref{fig:shortchain_f}(b) and \ref{fig:shortchain_N}(b), as the theoretical arguments reported in Section~\ref{sec:starblob} still hold in this extreme case.

\begin{figure}[!h]
\begin{center}
\includegraphics[width=0.65\textwidth]{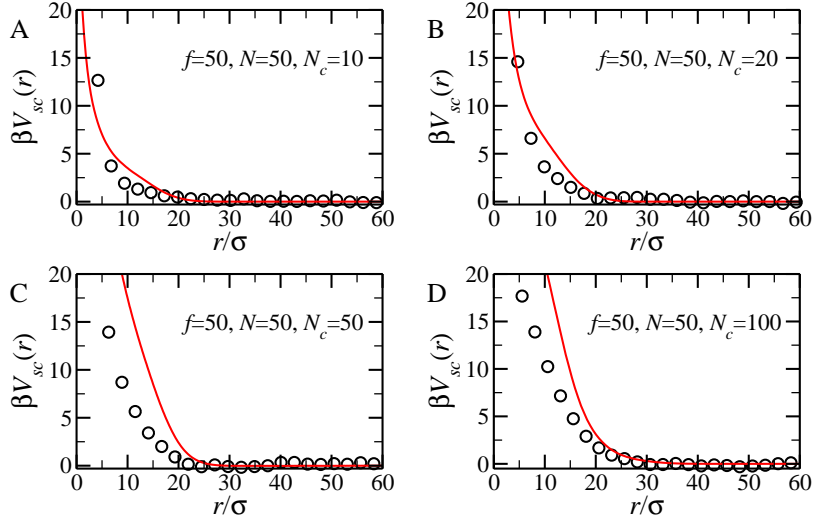}
\caption{Star polymer - linear chain interaction potential, for star polymers of fixed arm length $N$=50 and 
fixed functionality $f=50$, and linear chains of different length: (a)$N_c=10$, (b) $N_c=20$ (c) $N_c=50$, (d) $N_c=100$. 
Black dots refer to results obtained from monomer-resolved simulations, full red lines refer to results obtained from coarse-grained simulations.}
\label{fig:comp_cdm}
\end{center}
\end{figure}

Finally, we report the 
comparison of the star-chain effective potentials $V_{sc}(r)$, as done in Section~\ref{sec:comparison}.
In Fig.~\ref{fig:comp_cdm} we report the comparison between a star of functionality $f$=50 and arm length $N$=50 and linear chains of different lengths. 
It can be seen that the agreement between coarse-graining and monomer-resolved results is rather poor, especially for the two longer chains tested, $N_c=50$
and $N_c = 100$. In these cases, the coarse-graining vastly overestimates the interaction potentials. An explanation for this poor comparison can be formulated as follows.
The insertion of a monomer in the semidilute interior of the star brings forward a disturbance of the star profile, which extends over a range of the
local correlation length $\xi(r)$. The separation between two successive monomers in the chain is, of course, smaller than $\xi(r)$, so that the effect of introducing,
say, a dimer in the interior of the star is overestimated if we make the assumption of the superposition of the effects of the individual monomers. Once a monomer
has been introduced, it has created a disturbance of the star profile around it, and the second can be accommodated with less additional disturbance than 
what the superposition approximation assumes. By coarse-graining the star as a single object for the interaction with a succession of monomers, we are,
however, making precisely this approximation, ascribing to the star a stronger deformation than it actually has and thereby overestimating the effective
star-chain interaction.

When we consider a short linear chain of $N_0 = 10$ as the fundamental building block of a coarse-grained chain, 
two successive blobs are separated by a distance of the order $\xi(r)$ and thus
the overall rearrangement caused by two successive blobs will be very well approximated by the superposition approximation. In fact, both the 
star and the chain can rearrange to accommodate one into the other, and the energetic cost per monomer will be smaller with respect to the free monomers. From the perspective of 
the effective interaction, the final outcome will be a softer interaction potential, compared to superposition of star-monomer contributions, resulting from this approach. The 
crucial ingredients missing here are of course  correlations, arising from the organization of the monomers in the linear chain inside the same Daoud-Cotton blob. 

\bibliography{biblio}

\end{document}